\documentclass[aps,prd,twocolumn,showpacs,preprintnumbers,superscriptaddress,nofootinbib]{revtex4}
\usepackage{graphics}  
\usepackage{epsfig}
 \usepackage{verbatim}

\IfFileExists{srcltx.sty}{\usepackage[active]{srcltx}}

\newcommand{\be}{\begin{equation}}
\newcommand{\ee}{\end{equation}}
\newcommand{\bea}{\begin{eqnarray}}
\newcommand{\eea}{\end{eqnarray}}

\begin{document}

\title{Secondary photons and neutrinos from cosmic rays produced by distant blazars}

\author{Warren Essey}
\affiliation{Department of Physics and Astronomy, University of California, Los 
Angeles, CA 90095-1547, USA}

\author{Oleg E. Kalashev}
\affiliation{Institute for Nuclear Research, 60th October Anniversary Prospect 7a, Moscow 117312 Russia}

\author{Alexander Kusenko}
\affiliation{Department of Physics and Astronomy, University of California, Los 
Angeles, CA 90095-1547, USA}
\affiliation{IPMU, University of Tokyo, Kashiwa, Chiba 277-8568, Japan}

\author{John F. Beacom}
\affiliation{
Center for Cosmology and Astro-Particle Physics, Ohio State University, Columbus, Ohio 43210, USA}
\affiliation{Department of Physics, Ohio State University, Columbus, Ohio 43210, USA}
\affiliation{Department of Astronomy, Ohio State University, Columbus, Ohio 43210, USA}


\preprint{UCLA/09/TEP/61}

\begin{abstract}
Secondary photons and neutrinos produced in the interactions of cosmic ray protons emitted by distant Active Galactic Nuclei (AGN) with the photon background along the line of sight can reveal a wealth of new information about the intergalactic magnetic fields (IGMF), extragalactic background light (EBL), and the acceleration mechanisms of cosmic rays. The secondary photons may have already been observed by gamma-ray telescopes.  We show that the secondary neutrinos improve the prospects of discovering distant blazars by IceCube, and we discuss the ramifications for the cosmic backgrounds, magnetic fields, and AGN models. 

\end{abstract}

\maketitle


AGN are believed to be the most powerful sources of both $\gamma$ rays and cosmic rays.  The $\gamma$-ray observations are more easily associated with the sources, while the association of ultrahigh-energy cosmic rays (UHECR) with the sources is complicated by the deflections due to Milky Way magnetic fields~\cite{Harari:2000he}.  

It was recently pointed out that interactions of cosmic rays emitted by AGN with the photon background along the line of sight can produce $\gamma$ rays that may have already been observed by the Cherenkov telescopes~\cite{Essey:2009zg}.  The spectra of $\gamma$ rays observed from distant blazars~\cite{Aharonian:2005gh} are readily reproduced by the secondary photons produced in interactions of cosmic rays with the cosmic backgrounds~\cite{Essey:2009zg}.  While there is little doubt that AGN are ample sources of primary $\gamma$ rays, these primary photons are subject to attenuation at TeV energies due to the pair production losses on the extragalactic background light (EBL).  The secondary photons produced by proton-photon interactions nearby can replace the primary photons in the high-energy tails of the spectra observed from the most distant blazars.   

This possibility, which is interesting in its own right, has far-reaching implications for both the extragalactic background light (EBL) and intergalactic magnetic fields (IGMF).  Indeed, $\gamma$-ray observations provide a unique probe of  EBL, assuming the signals are pure primary photons, uncontaminated by secondary photons~\cite{Protheroe:2000hp,Aharonian:2005gh,Stecker:2007zj,Mazin:2007pn,Stecker:2007jq,Franceschini:2008tp,Gilmore:2009zb}.    This assumption may be incorrect since protons emitted by AGN can contribute to $\gamma$-ray signals, as evidenced 
by the $\gamma$-ray spectra of several distant blazars that do not show an expected attenuation at energies above TeV.  The lack of attenuation could be evidence of a relatively low EBL~\cite{Stecker:2005qs,Stecker:2007jq,Stecker:2008fp,Gilmore:2009zb}, very hard emission spectra~\cite{Aharonian:2005gh}, or of some new physics in the form of axion-like particles~\cite{De_Angelis:2007dy}, or Lorentz invariance violation~\cite{Protheroe:2000hp}.   A less exotic explanation is that the secondary photons replace primary photons in the high-energy tails of the AGN spectra, hence creating a spectrum without a cutoff that is well fit to AGN at relatively high redshift ~\cite{Essey:2009zg}.

In this paper we explore the multi-messenger signals of AGN focusing, in particular, on the neutrinos accompanying the secondary photons.   AGN are expected to accelerate cosmic rays to energies up to $\sim 10^{11}$ GeV, but most likely have a high energy cutoff below that~\cite{1995ApJ...454...60N}  due to interactions at the source. Thus we will consider cutoffs in the range $10^8$ GeV - $10^{11}$ GeV. Cosmic rays with energies below the Greisen-Zatsepin-Kuzmin (GZK) cutoff~\cite{gzk} of about $3\times 10^{10}$~GeV can cross cosmological distances without a significant energy loss.  However, with a small probability, these protons do interact with the cosmic backgrounds and produce photons.  Our investigation differs form the earlier studies of neutrino signals from AGN~\cite{Stecker:1991vm}, which assumed that all the pion production occurs at the source.  We will discuss the important differences in the two types of signals: from the hadronic interactions \textit{in situ} and along the line of sight.    We will concentrate on protons with energies below the GZK cutoff, which can travel cosmological distances; we will not discuss the cosmogenic neutrinos produced by protons with  energies above the GZK cutoff~\cite{Berezinsky:1970xj}.

The secondary photons are generated in two types of interactions of UHECR along the line of sight. First, the proton interactions with CMB photons can produce electron-positron pairs and give rise to an  electromagnetic cascades due to what is called proton pair production (PPP), $p \gamma_{_{\rm CMB}}\rightarrow pe^+e^-$~\cite{Blumenthal:1970nn}.  Second,  proton interactions with the EBL can produce pions, which decay and produce photons as well in the reactions $p \gamma_{_{\rm EBL}} \rightarrow p\pi^0$ or $p \gamma_{_{\rm EBL}} \rightarrow n\pi^+ $.  While the PPP process is not associated with any neutrinos, the pion photoproduction generates a neutrino flux related to the $\gamma$-ray flux.  The relative importance of the two processes depends on the proton injection spectrum, which we will parameterize by a constant power-law exponent $\alpha$ and  maximal energy  $E_{\rm max}$:
\begin{equation} 
 F(E)\propto E^{-\alpha} \ \exp (E_{\rm max}-E).
\end{equation}
Although the spectral index $\alpha=2.7^{+0.05}_{-0.15} $ gives a good fit to the UHECR data at the highest energies~\cite{Berezinsky:2002nc}, the measured spectrum is a superposition of individual sources with different values of $E_{\rm max}$.  Thus we consider a smaller value of $\alpha \approx2$, which agrees with the data at lower energies~\cite{Berezinsky:2002nc}. The parameters $\alpha$ and  $E_{\rm max}$ determine the power in the highest-energy cosmic rays, which pile up around the GZK energy and contribute to the PPP process.  As for the pion photoproduction on EBL, it is mainly due to the lower part of the proton spectrum, at energies of the order of $10^8$~GeV.   The predicted spectral shape of secondary photons is not very sensitive to the variations in $\alpha$ and  $E_{\rm max}$;  it is determined primarily by the spectrum of the background photons.  The model predictions agree with the data on the most distant sources~\cite{Essey:2009zg} for an effective luminosity of a single AGN in cosmic rays above $10^7$~GeV in the range $L_{\rm eff}= (10^{47}-10^{49})\, {\rm erg/s}$. The observed luminosity is boosted by the beaming factor due to the AGN's relativistic velocity with the blazar jet pointing at Earth. Thus we can express the effective luminosity as
\begin{equation}
L_{\rm eff} \sim 10^2 f_{\rm p} \times \left(\frac{f_{\rm beam}}{100}\right)  L_{\rm source},
\end{equation}
where $f_{\rm p} \lesssim 1$ is the fraction of protons in cosmic rays (which may also contain some heavy nuclei~\cite{PhysRevLett.71.3401}). Thus a source luminosity of $10^{45}-10^{47} $ erg/s could easily provide the required power in protons and is consistent with theoretical models~\cite{Berezinsky:2002nc}.

While the AGN energetics and fluxes are consistent with observations of secondary photons, and while the spectra observed by MAGIC and VERITAS are well fit by the model~\cite{Essey:2009zg}, the secondary photons only point back to the sources if the IGMF is  relatively small.  Since the $p \gamma_{_{\rm EBL}}$ interactions take place well outside the galaxy clusters of both the source and the observer, the cluster magnetic fields are irrelevant for this discussion.  
Although one expects larger fields in filaments and walls, only the IGMF present deep in the voids along the line of sight is important~\cite{Essey:2009zg}.  Within the host galaxy, the directions of the protons could be altered by the galactic magnetic fields, but the broadening of the image due to any deflections in the host galaxy cannot exceed $\Delta \theta \sim r/D_{\rm source}$, where $r$ is the size of the host galaxy, and $D_{\rm source}$ is the distance to it.  Furthermore, the outflows of the jets from an AGN are likely to contain coherent magnetic fields aligned with the jet, so that the accelerated protons remain in the scope of the initial jet rather than get deflected, so that the beaming factors are not affected significantly. The possible thin walls of magnetic fields that may intersect the line of sight cannot introduce a deflection by more than $\Delta \theta\sim h/D_{\rm wall} $, where $h$ is the wall thickness and $D_{\rm wall} $ is the distance to the wall.    

However, the IGMF in the voids can defocus the images of distant sources in secondary photons. Constrained simulations~\cite{2004JETPL..79..583D,Sigl:2004yk,Das:2008vb} predict a very inhomogeneous distribution of the magnetic fields characterized by large voids, thin voids, and thin filaments, but, as emphasized in Ref.~\cite{2004JETPL..79..583D}, the results of these models should be taken as upper limits on IGMF.  While IGMF with nano-Gauss strengths can be accommodated in a model,  the fields in the voids as low as $10^{-18}$~G are also consistent with all the astrophysical data. 
If the galactic magnetic fields arise as a result of a dynamo mechanism, the  
seed fields of the order of $10^{-20}$~G are required.  The seed fields can be primordial~\cite{Grasso:2000wj}, or they can originate from the Biermann effect~\cite{biermann}. Future studies of gamma ray signals can detect or constrain the IGMF~\cite{Plaga,Essey:2009zg}.  

The secondary photons preserve the directions and create  images of distant blazars if the average IGMF in the voids are of the order of $10^{-14}$~G or smaller\cite{Essey:2009zg}.  One important test is the time variability.  A rapid time variability can be destroyed by the delays in the proton arrival times caused by the IGMF.  On the other hand, such delays can explain the reported mismatch of time scales for X-ray and VHE variabilities~\cite{Harris:2009wn}, as well as some unusual energy-dependent time lag~\cite{Albert:2007qk}.  One has to be careful drawing conclusions from these observations because photons of different energies may originate from different regions in the source, and also because one must distinguish between the stochastic fluctuations in the arrival times of secondary photons and an intrinsic source variability.  We leave this for future work. 

Aside from some possible exotic explanations, one can phrase the dilemma as follows: the lack of high-energy attenuation in the spectra of distant $\gamma$-ray sources indicates either a low EBL, or a low IGMF.  In the former case, the observed $\gamma$ rays can come from distant sources (except for those cases where the local photon density may be too  high to allow the escape of TeV photons~\cite{2008AIPC.1085..644C,2008ApJ...685L..23A}).  In the latter case, the observed $\gamma$ rays are secondary photons produced along the line of sight.  Of course, if both EBL and IGMF are small, then the observed signal can be a combination of primary and  secondary photons, and the ratio of secondary to primary fluxes should increase with the distance to the source.  

If IceCube observes AGN as sources of UHE neutrinos, one can confirm that hadronic interactions do take place, and there should be a corresponding flux of photons.  If the images of AGN in photons and neutrinos are not pointlike, but are surrounded by halos, it is a confirmation that the pion photoproduction takes place along the line of sight, not at the source.   Of course, the angular resolution of IceCube limits the application of this test. 

We show in Fig.~1 the neutrino spectra calculated numerically for a single AGN at redshift z=0.14, such as 1ES0229+200, for which the secondary photons fit HESS observations very well (Fig.~2).  We have also considered neutrinos from a closer source, such as Mrk421 or Mrk501, for which the gamma-ray spectra probably contain a mixture of primary and secondary photons, with the latter dominating at high energies. 
%
\begin{figure}[h!]
  \begin{center}
      \includegraphics[width=80mm]{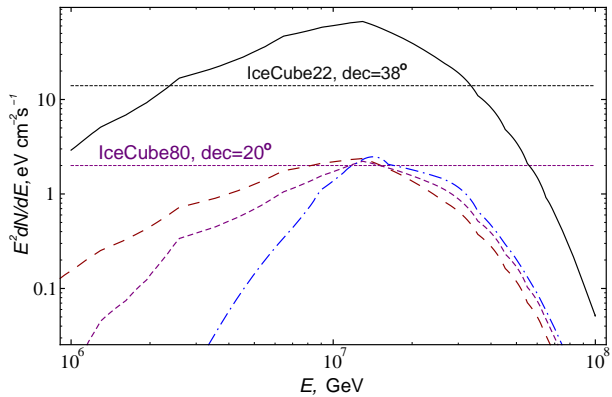}
\caption{Expected neutrino spectra from an AGN at  $z=0.14$ (such as 1ES0229+200) for three EBL models:  Franceschini \textit{et al.}~\cite{Franceschini:2008tp} (dash-dotted line), Primack \textit{et al.}~\cite{Primack:2008nw} (short-dashed line), and Stecker \textit{et al.}~\cite{Stecker:2005qs} (long-dashed line). The cosmic ray emission power above $10^7$~GeV, $L_{\rm source}$, consistent with HESS data for $f_{\rm beam}=10^2$ and $E_{\rm max}=10^8$GeV is $2\times 10^{47}$erg/s, $1.5\times 10^{47}$erg/s, and $6\times 10^{46}$erg/s, respectively.  The solid line corresponds to an AGN at $z=0.03$, such as Mrk421 and Mrk501, assuming $L_{\rm source}= 5\times 10^{47}$erg/s and $f_{\rm beam}=10^2$. The IceCube sensitivity for 1ES0229+200 is shown for 80 strings with 1 year exposure time, and the sensitivity for Mrk421 and Mrk501 is for 22 strings with 0.75 year exposure time~\cite{2009arXiv0910.3644D}.}
    \end{center}
  \label{fig:3model}
\end{figure}

IceCube can search for point sources in up-going events at low energy~\cite{Lauer:2009yv}, and can explore the higher energies using the downgoing events~\cite{Abbasi:2009cv}. The results presented in Fig.~1 are not within the current reach of IceCube, but should be within the reach of IceCube80 for the Stecker EBL model. Of course the lower EBL models would require a longer exposure time to be within IceCube80's sensitivity.

The AGN models have many uncertainties, and a deviation in the model parameters from the values we have assumed for the spectrum and the beaming factor of UHECR can affect the predictions of our model, but a variety of parameters leave the AGN within IceCube's predicted range. For instance $\alpha>2$ increases the $L_{\rm eff}$ needed to normalize to gamma ray data and makes the source brighter for IceCube. Thus IceCube may be able to distinguish between certain model parameters. In particular, the position of the energy peak in neutrinos is sensitive to $E_{\rm max}$, unlike the gamma rays. 
\begin{figure}[t!]
  \begin{center}
      \includegraphics[width=60mm,angle=270]{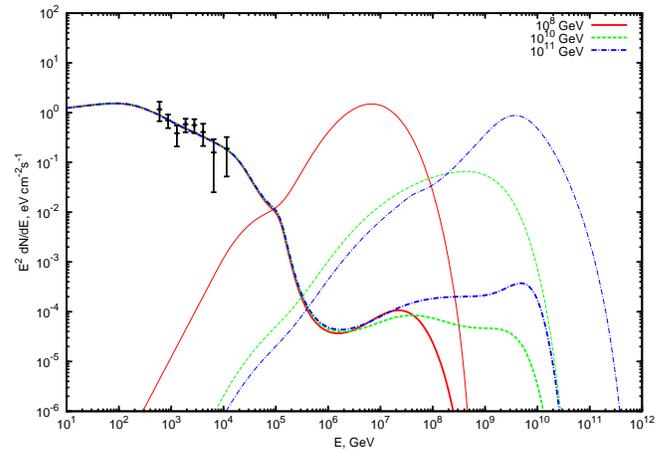}
\caption{Photon (low energy) and neutrino (high energy) spectra expected from an AGN at  $z=0.14$ (such as 1ES0229+200), normalized to HESS data points (shown), with $\alpha=2 $,  for EBL of Ref.~\cite{Stecker:2005qs}, and for  $E_{\rm max}=10^8$GeV, $10^{10}$GeV, and $10^{11}$GeV shown by the solid, dashed, and dash-dotted lines, respectively. The $L_{\rm eff}$ for each $E_{\rm max}$ was set at $6\times 10^{48}$erg/s, $2\times 10^{46}$erg/s, and $9\times 10^{45}$erg/s, respectively.}
    \end{center}
  \label{fig:alpha2}
\end{figure}

If the point sources are detected by IceCube, one can test the hypothesis of secondary photons and neutrinos.  One test is the fuzziness of the pointlike sources: equal halos around the blazar images in $\gamma$ rays and neutrinos would indicate that both of them are secondary particles produced in proton interactions with EBL. The application of this test may be complicated by the limitations in the IceCube angular resolution. However, the mere fact that some very distant sources are observed, along with some luminosity-distance relations can prove that the observed particles are secondary. The proton flux is proportional to $1/D_{\rm source}^2$, while the probability of a proton interaction with EBL is proportional to distance $D_{\rm source}^{1+\delta (z)}$.  The parameter $\delta (z)$ accounts for the evolution of the EBL with redshift and varies over the range $\delta (z)= 0.2 - 0.8$ for redshifts $z = 0.1 -1$.  The resulting $1/D_{\rm source}^{1-\delta (z)}$ scaling of both secondary photons and neutrinos could be used as a statistical test of their origin by comparing the signals from blazars at different distances.  This is different from the $1/D_{\rm source}^2$ scaling expected for primary neutrinos produced at the source, and it is also different from the expected spectral properties of $\gamma$ rays, which should show a suppression due to their interaction with EBL.  This is probably the strongest test of the model, since such a scaling would be difficult to explain otherwise. A detection of some distant sources by IceCube would help distinguish between our model and the alternatives since a $1/D_{\rm source}^2$ scaling would quickly drop the flux below IceCube's sensitivity.  The data on the variety of distant AGN observed by the existing $\gamma$-ray telescopes may also allow an application of this test in the very near future. 

If our mechanism, along with its requisite assumption of low IGMF, is confirmed, then $\gamma$-ray and neutrino data may be used to study the UHECR sources at distances far beyond the GZK cutoff.  For example, $\gamma$-ray bursts (GRB) are also likely sources of UHECR~\cite{Waxman:1995vg}.  The GRB observed by the $\gamma$-ray instruments occur at distances well in excess of the GZK radius.  However, the interactions of UHECR produced in GRB with EBL along the line of sight can generate the fluxes of photons and neutrinos that can be used to confirm the production of UHECR in GRB.  

The unique properties of our model mentioned above should be easily testable in the near future since the multi-wavelength and multi-particle observations of AGN are ongoing.  With more data, it will be possible to test scaling arguments, particle content,  and the absence of the spectral suppression features, so that one can distinguish  between the models and discover some fundamental properties of both AGN and the intergalactic space.  The neutrino observations are likely to play a key role in this exploration. 

We note in passing that our interpretation of $\gamma$-ray observations broadens the range of possibilities for hadronic models of $\gamma$-ray production, such as in Ref.~\cite{Ribordy:2009jj}, and allows the primary neutrinos to have a different spectrum.  So far, the hadronic models~\cite{Ribordy:2009jj} concentrated on $pp$ reactions, and not $p\gamma$ reactions, because the high density of photons at the source would stymie the production of primary very high energy (VHE) $\gamma$ rays.   This constraint is eliminated by the possibility that it is secondary, not primary photons that account for the observed VHE spectra of blazars. 
Since, in our case, the secondary TeV $\gamma$ rays are produced outside of the host galaxy, their observation is not in contradiction with the assumption of a high photon density at the source, which is required for neutrino production in $p\gamma$ interactions.  We leave this possibility for future work.

The authors thank T.~Arlen, I.~Tkachev, and V.~Vassiliev for discussions.  The work of W.E. and A.K. was supported  by DOE grant DE-FG03-91ER40662 and NASA ATFP grant  NNX08AL48G.  The work of O.K. was supported by RFBR grant 07-02-00820. J.F.B. was supported by NSF CAREER Grant PHY-0547102. 


\end{document}